\documentclass[aps,prl,twocolumn,floats]{revtex4}

\usepackage{ifthen}
\usepackage{ifpdf}

\ifpdf
\usepackage{graphicx}
\usepackage{epstopdf}
\else
\usepackage{graphicx}
\usepackage{epsfig}
\fi

\usepackage{times}
\usepackage{float}
\usepackage{amsmath}
\usepackage{amssymb}
\usepackage{bm}
\usepackage{latexsym}

\begin{document}
\newcommand{\hide}[1]{}
\newcommand{\tbox}[1]{\mbox{\tiny #1}}
\newcommand{\half}{\mbox{\small $\frac{1}{2}$}}
\newcommand{\sinc}{\mbox{sinc}}
\newcommand{\const}{\mbox{const}}
\newcommand{\trc}{\mbox{trace}}
\newcommand{\intt}{\int\!\!\!\!\int }
\newcommand{\ointt}{\int\!\!\!\!\int\!\!\!\!\!\circ\ }
\newcommand{\eexp}{\mbox{e}^}
\newcommand{\bra}{\left\langle}
\newcommand{\ket}{\right\rangle}
\newcommand{\EPS} {\mbox{\LARGE $\epsilon$}}
\newcommand{\ar}{\mathsf r}
\newcommand{\im}{\mbox{Im}}
\newcommand{\re}{\mbox{Re}}
\newcommand{\bmsf}[1]{\bm{\mathsf{#1}}}
\newcommand{\mpg}[3][b]{\begin{minipage}[#1]{#2}{#3}\end{minipage}}

\title{ The twilight zone in the parametric evolution 
of eigenstates: beyond perturbation theory and semiclassics}

\author{\small
J. A. M\'endez-Berm\'udez$^{1}$,
Tsampikos Kottos$^{1}$,
Doron Cohen$^{2}$
}

\affiliation{
$^{1}$ Max-Planck-Institut f\"ur Dynamik und Selbstorganisation, \\
Bunsenstra\ss e 10, D-37073 G\"ottingen, Germany
$^{2}$ Department of Physics, Ben-Gurion University, Beer-Sheva 84105, Israel 
}



\begin{abstract}

Considering a quantized chaotic system, 
we analyze the evolution of its eigenstates 
as a result of varying a control parameter. 
As the induced perturbation becomes larger,  
there is a crossover from a perturbative 
to a non-perturbative regime, which is reflected 
in the structural changes 
of the local density of states.
For the first time the {\em full} scenario 
is explored for a physical system: 
an Aharonov-Bohm cylindrical billiard. 
As we vary the magnetic flux,  
we discover an intermediate twilight regime 
where perturbative and semiclassical features co-exist.
This is in contrast with the {\em simple} crossover 
from a Lorentzian to a semicircle line-shape 
which is found in random-matrix models. 

\end{abstract}

\maketitle


The analysis of the evolution of eigenvalues and of the structural changes that the 
corresponding eigenstates of a chaotic system exhibit as one varies a parameter $\phi$ 
of the Hamiltonian ${\cal H}(\phi)$ has sparked a great deal of research activity for 
many years. Physically the change of $\phi$ may represent the effect of some externally 
controlled field (like electric field, magnetic flux, gate voltage) or a change of an 
effective-interaction (as in molecular dynamics). Thus, these studies are relevant for 
diverse areas of physics ranging from nuclear \cite{HZB95,W55} and atomic physics 
\cite{TAA95,FGGP99} to quantum chaos \cite{W88,CK01,MLI04,B03} and mesoscopics 
\cite{VLG02,TA93}.

Up to now the majority of this research activity was focused 
on the study of eigenvalues, where a good understanding has been 
achieved, while much less is known about eigenstates.
The pioneering work in this field has been done by Wigner \cite{W55}, 
who studied the parametric evolution of eigenstates 
of a simplified Random Matrix Theory (RMT) 
model of the type ${\cal H}=\bm{E}+\phi \bm{B}$. 
The elements of the diagonal matrix $\bm{E}$ are the 
ordered energies $\{E_n\}$, with mean level spacing $\Delta$, 
while $\bm{B}$ is a banded {\it random} matrix. 
Wigner found that as the parameter $\phi$ increases the 
eigenstates undergoes a transition from 
a perturbative {\em Lorentzian-type line shape} 
to a non-perturbative {\em semicircle line-shape}.

For many years the study of parametric evolution for {\it canonically quantized systems} 
was restricted to the exploration of the crossover from integrability to chaos \cite{MLI04,B03}. 
Only later \cite{CK01} it has been realized that a theory is lacking for systems 
that are chaotic to begin with. 
Inspired by Wigner theory, the natural prediction was that the local density of 
states (LDOS) should exhibit a crossover from a regime where a perturbative 
treatment is applicable, to a regime where semiclassical approximation is valid. However, 
despite a considerable amount of numerical efforts \cite{CK01}, there was no clear-cut 
demonstration of this crossover. Neither a theory has been developed describing how the 
transition from the perturbative to the non-perturbative regime takes place.


It is the purpose of this Letter to present, for the first time, 
a complete scenario of parametric evolution, 
in case of a physical system that exhibits {\it hard} chaos. 
We explore the validity of perturbation theory and semiclassics,  
and we discover the appearance of an intermediate regime 
(``twilight zone'') where {\it both perturbative and semiclassical 
features co-exist}.  Without loss of generality we consider 
as an example a billiard system whose classical dynamics 
is characterized by a correlation time $\tau_{\tbox{cl}}$,   
which is simply the ballistic time. 
Associated with $\tau_{\tbox{cl}}$ is the energy scale 
$\hbar/\tau_{\tbox{cl}}$. Next we look on a {\em similar} billiard, 
but with a rough boundary. This roughness is characterized by 
a length scale which is $\ell$ times smaller, 
hence we can associate with it an energy scale  
$\delta E_{\tbox{NU}} = (\hbar/\tau_{\tbox{cl}})\times \ell$. 
The roughness does not affect the chaoticity:  
the correlation time $\tau_{\tbox{cl}}$ 
as well as the whole power spectrum are barely affected. 
Consequently we explain that $\delta E_{\tbox{NU}}$ 
is not reflected in the RMT modeling of the Hamiltonian.
Still in the LDOS analysis we find that 
non-universal (system specific) features appear.  
The appearance of such features is a {\em generic} phenomenon 
in quantum chaos studies. It introduces a new ingredient into 
the theory of parametric evolution {\em which goes beyond RMT}.

\hide{
The resulting structures can be understood using 
a phase-space picture, and are related to the non-semiclassical 
features of the wavefunctions. In other words, on the one hand 
we demonstrate the limitations of random matrix modeling, 
while on the other hand we show the consequences of having failure 
of the so-called ``Berry conjecture" \cite{berry}.  
}

\begin{figure}[b]
\includegraphics[clip,width=0.8\hsize]{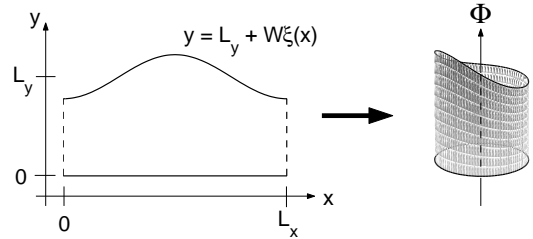}
\caption{
Left: 2D billiard with $\ell=1$. Right: Corresponding Aharonov-Bohm cylinder.} 
\end{figure}


The model that we will use in our analysis 
is a particle confined to an Aharonov-Bohm (AB)
cylindrical billiard (see Fig.1) where one can control 
the magnetic flux $\Phi$.
The cylindrical billiard is constructed by wrapping 
a 2D billiard with hard wall boundaries.
The lower boundary at $y=0$ is flat, while the 
upper boundary $y=L_y+W\xi(x)$ is deformed. 
The deformation is described by $\xi(x)=\sum_{n=1}^{\ell} A_n \cos(nx)$
where $A_n$ are random numbers in the range [-1,1].
The illustration in Fig.1 assumes a smooth boundary ($\ell=1$).
The Hamiltonian of a particle in the cylindrical AB billiard is
\begin{equation}
{\cal H}(\phi) = 
\frac{1}{2\mathsf{m}} 
\left[ \left( p_x - \frac{e}{L_x}\Phi \right)^2 + p_y^2 \right]
\label{Hcl}
\end{equation}
supplemented by $L_x$ periodic boundary conditions
in the horizontal direction, and hard wall boundary
conditions along the lower and upper boundaries.
$p_x$ and $p_y$ are the momenta. Later we shall use
the notation $\phi=e\Phi/\hbar$.
We consider the chaotic ${\cal H}(\phi=0)$ 
as the unperturbed Hamiltonian.


After conformal transformation \cite{MLI04} the billiard
is mapped into a rectangular, with a mass tensor which is space
dependent. Then it is possible to compute the matrix representation
of the Hamiltonian in the plane wave basis $|\nu\mu\rangle$
of the rectangular. The result is:
\begin{eqnarray} 
&& \hspace*{-0.5cm}
{\cal H}_{\nu\mu,\nu'\mu'}(\phi) =
\frac{\hbar^2}{2\pi \mathsf{m}}
\left\{
\pi \left( \nu-\frac{\phi}{2\pi} \right) ^2 \delta_{\nu,\nu'}
\delta_{\mu,\mu'}
\right.
\nonumber \\  
&& +
\left[
\frac{\mu^2}{8\alpha^2}J^{(0,2)}_{\nu'\nu} +
\epsilon^2 J^{(2,2)}_{\nu'\nu}
(\frac{1}{8}{+}\frac{\pi^2 \mu^2 }{6}) \right] \delta_{\mu,\mu'}
+ (-1)^{\mu{+}\mu'}\mu \mu'
\nonumber \\
&& \times
\left.
\left[
\epsilon^2\frac{2(\mu^2+\mu'^2)}
{(\mu^2-\mu'^2)^2}J^{(2,2)}_{\nu'\nu}
-i\epsilon
\frac{\nu+\nu'-\phi/\pi} {\mu^2-\mu'^2} J^{(1,1)}_{\nu'\nu}
\right]
\right\}
\label{hmn}
\end{eqnarray}
where
\begin{eqnarray} \nonumber
J^{(l,k)}_{\nu'\nu} =
\int^{L_x}_0 dx \, \eexp{i(\nu'-\nu)2\pi x/L_x}
\left(\frac{d\xi}{dx}\right)^l
\frac{1}{(1+\epsilon\xi(x))^k}
\end{eqnarray}
The classical dimensionless parameters
of the model are the aspect ratio $\alpha=L_y/L_x$,
the tilt relative amplitude $\epsilon=W/L_y$,
and the roughness parameter $\ell$.
Upon quantization we have $\hbar$ that together with $\mathsf{m}$ and $E$
determines the De-Broglie wavelength of the particle,
and hence leads to an additional dimensionless parameter
$n_E=[L_xL_y/(2\pi\hbar^2)]\mathsf{m}E$.
For 2D billiards the mean level spacing $\Delta$ is constant,
and hence $n_E=E/\Delta \propto 1/\hbar^2 $
can be interpreted as either the scaled energy
or as the level index. Optionally we define
a semiclassical parameter $\hbar_{\tbox{scaled}}=1/\sqrt{n_E}$.


In the numerical study we have taken $\epsilon{=}0.06$ 
and $\alpha{=}1$, for which the classical dynamics  
is completely chaotic (for any~$\phi$). 
We consider either $\ell{=}1$ for smooth boundary,  
or $\ell{=}100$ for rough boundary.  
The eigenstates $|n(\phi)\rangle$ of the Hamiltonian ${\cal H}(\phi)$ 
were found numerically for various values of 
the flux (\mbox{$0.0006 < \phi < 60$}). 
We were interested in the states within 
an energy window $\delta E {\approx} 45$ that contains $\delta n_E {\sim} 200$ levels 
around the energy $E {\approx} 400 $. Note that the size 
of the energy window is classically small (\mbox{$\delta E \ll E$}),  
but quantum mechanically large (\mbox{$\delta E \gg \Delta$}).


The object of our interest are the overlaps 
of the eigenstates $|n(\phi)\rangle$ 
with a given eigenstate $|m(0)\rangle$ 
of the unperturbed Hamiltonian:
\begin{eqnarray} \label{e3}
\hspace*{-0.5cm}
P(n|m) = |\langle n(\phi)|m(0)\rangle|^2 
= \int \frac{dx dy dp_x dp_y}{(2\pi\hbar)^2} \rho^{(n)}\rho^{(m)} 
\end{eqnarray}
The overlaps $P(n|m)$ can be regarded as a distribution 
with respect to $n$. Up to some trivial scaling it
is essentially the local density of states (LDOS). 
The associated dispersion is defined as  
$\delta E = [\sum P(n|m) (E_n-E_m)^2]^{1/2}$
In practice we plot $P(n|m)$ as a function 
of $r=n-m$ or as a function of $(E_n{-}E_m)$, 
and average over the reference state $m$. 
The second equality in (\ref{e3}) is useful 
for the semiclassical analysis. It involves 
the Wigner functions $\rho^{(n)}(x,y,p_x,p_y)$ 
which are associated with 
the eigenstates $|n(\phi)\rangle$. 
The semiclassical approximation is based 
on the microcanonical approximation 
$\rho^{(n)} \propto \delta(E_n{-}{\cal H}(x,y,p_x,p_y))$. 
With this approximation the integral 
can be calculated analytically leading to  
\begin{eqnarray}
P_{\tbox{cl}}(n|m) = 
\frac{\Delta}
{\pi\sqrt{2(\delta E_{\tbox{cl}})^2 - 
\left[ (E_n{-}E_m) - \delta E_{\tbox{cl}}^2/(2E_m) \right]^2}}
\label{Pcl}
\end{eqnarray}
where $\delta E_{\tbox{cl}} = (\hbar v_E / L_x)\phi$ 
with $v_E = (2E/\mathsf{m})^{1/2}$. 
It is implicit in (\ref{Pcl}) 
that $P_{\tbox{cl}}(n|m){=}0$ outside of the allowed range,  
which is where the expression under the square~root
is negative: For large $|E_n{-}E_m|$ there 
is no intersection of the corresponding energy 
surfaces, and hence no classical overlap.


A few words are in order regarding quantum 
to classical correspondence (QCC).  
Whenever $P(n|m)\approx P_{\tbox{cl}}(n|m)$  
we call it ``detailed QCC", 
while $\delta E \approx \delta E_{\tbox{cl}}$
is referred to as ``restricted QCC" \cite{CK01}. 
It is remarkable that (the robust) 
restricted QCC holds even if 
(the fragile) detailed QCC fails completely. 
We have verified \cite{MCK04}
that also in the present system  $\delta E$ 
is numerically indistinguishable from $\delta E_{\tbox{cl}}$.

 
A fixed assumption of this work is that $\phi$ 
is classically small. But quantum mechanically it 
can be either `small' or `large'.  
Quantum mechanically {\em small} $\phi$ means  
that perturbation theory do provide 
a valid approximation for $P(n|m)$. 
What is the {\em border} between the perturbative regime 
and the non-perturbative regime, we discuss later. 
First we would like to show that the prediction which 
is based on perturbation theory, 
to be denoted as $P_{\tbox{prt}}(n|m)$, 
is very different from the semiclassical approximation.

In order to write the expression for $P_{\tbox{prt}}(n|m)$
we have first to clarify how to apply perturbation 
theory in the context of the present model. 
To this end, we write the perturbed Hamiltonian ${\cal H}(\phi)$ 
in the basis of ${\cal H}(\phi=0)$. 
Since we assume that the perturbation 
is classically small, it follows that we can linearize 
the Hamiltonian with respect to $\phi$. 
Consequently the perturbed Hamiltonian is written 
as ${\cal H} = \bm{E} + \phi \bm{B}$, 
where $\bm{E}=\mbox{diag}\{E_n\}$ is a diagonal matrix, 
while $\bm{B} = \{ -(\hbar/e){\cal I}_{nm} \}$.
The current operator is conventionally defined as
\begin{eqnarray} \nonumber
{\cal I}\equiv -{\partial {\cal H}/\partial \Phi} = (e/(\mathsf{m}L_x))p_x
\end{eqnarray}
Its matrix elements can be found using a semiclassical recipe \cite{FP86},
namely $|{\cal I}_{nm}|^2 \approx (\Delta /(2 \pi\hbar)) \tilde{C}((E_n{-}E_m)/\hbar)$,
where $\tilde{C}(\omega)$ is the Fourier transform
of the current-current correlation function $C(\tau)$.
Conventional condensed matter calculations are done
for {\em disordered} rings where one assumes
$C(\tau)$ to be exponential, with time constant $\tau_{\tbox{cl}}$
which is essentially the ballistic time.
Hence $\tilde{C}(\omega) \propto 1/(\omega^2+(1/\tau_{\tbox{cl}})^2)$
is a Lorentzian. This Lorentzian approximation
works well also for the chaotic ring that we consider.
In fact we can do better by exploiting a relation
between ${\cal I}(t)$ and the force ${\cal F}(t)=-{\dot p_x}$,
leading to $\tilde{C}(\omega) = (e/(\mathsf{m}L_x))^2 
{\tilde C}_{\cal F}(\omega)/\omega^2$.
The force ${\cal F}(t)$ is a train of spikes corresponding
to collisions with the boundaries. Assuming that the collisions 
are uncorrelated on short times we have
${\tilde C}_{ \cal F} (\omega) \approx (8/3\pi)\mathsf{m}^2v_E^3/L_y$,
for $\omega \gg (1/\tau_{\tbox{cl}})$.  
This is known as the ``white noise" approximation \cite{BC}.
We have checked the validity of this approximation
in the present context by a direct numerical evaluation
of ${\tilde C}(\omega)$, and also verified the validity of 
the above recipe by direct evaluation of the matrix elements of 
$\bm{B}$ via Eq.(\ref{hmn}), see Fig.~2(a). 
The classical ${\tilde C}(\omega)$ was numerically evaluated 
by Fourier analysis of the fluctuating current $\mathcal{I}(t)$ 
for a very long ergodic trajectory that covers densely 
the whole energy surface ${\cal H}(0)=E$.

\begin{figure}
\includegraphics[clip,width=1.0\hsize]{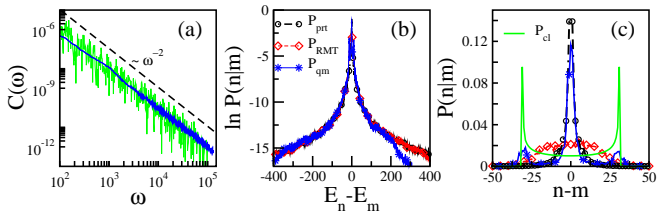}
\caption{(a) The classical power spectrum $C(\omega)$ plotted (in grey) 
together with the quantum mechanical 
band-profile $(2\pi e^2/\Delta \hbar)|{\bf B}_{nm}|^2$ 
for $\ell=1$, and $\hbar_{\rm scaled}\approx 0.018$. 
(b) The LDOS kernel $P(n|m)$ in the perturbative regime 
for a billiard with $\ell=100$ and perturbation $\phi=2.7$. 
(c) Same as (b) but zoomed normal scale. 
The width of the non-perturbative component is $\Gamma/\Delta=36$. 
Note that in this regime the variance $\Delta E/\Delta \approx 58$ 
is still dominated by the (perturbative) tails. For comparison we display the 
calculated $P_{\tbox{prt}}$, $P_{\tbox{cl}}$, and $P_{\tbox{RMT}}$.}
\end{figure}

Perturbation theory to infinite order
with the Hamiltonian
${\cal H} = \bm{E} + \phi \bm{B}$
leads to a Lorentzian-type approximation
for the LDOS \cite{W55} 
(see also Section 18 of \cite{CK01}c). 
It is an approximation because
all the higher orders are treated within
a Markovian-like approach (by iterating
the first order result) and convergence
of the expansion is pre-assumed, leading to
$P_{\tbox{prt}}(n|m) =
\phi^2 |\bm{B}_{nm}|^2 / [\Gamma^2 + (E_n{-}E_m)^2]$.
In practice the parameter $\Gamma(\phi)$ can be determined
(for a given $\phi$) by imposing the requirement of
having $P_{\tbox{prt}}(r)$ normalized to unity.
Substituting the expression for the matrix elements we get
\begin{eqnarray}
\label{pert}
P_{\tbox{prt}}(n|m) =
\frac{8\hbar^2(\hbar v_E)^3/(3\pi \mathsf{m}L_y^2 L_x^3)}
{(E_n{-}E_m)^2+(\hbar/\tau_{\tbox{cl}})^2}
\frac{\phi^2}{(E_n{-}E_m)^2+\Gamma^2}
\end{eqnarray}
By comparing the exact $P(r)$ to the approximation Eq.(\ref{pert})
we can determine the regime $\phi < \phi_{\tbox{prt}}$ 
for which the approximation $P(r) \approx P_{\tbox{prt}}(r)$ makes sense.
The practical procedure to determine $\phi_{\tbox{prt}}$
is to plot $\delta E_{\tbox{prt}}$ and to see where 
it departs from $\delta E_{\tbox{cl}}$. 
The latter is a linear function of $\phi$ 
while the former becomes sublinear for large enough $\phi$, 
(and even would exhibit saturation if we had a finite bandwidth). 
In case of Eq.(\ref{pert}) this reasoning leads 
to a crossover when ${\delta E_{\tbox{cl}}(\phi) \sim \hbar/\tau_{\tbox{cl}}}$. 
Hence we get that the border of the perturbative regime 
(see footnote \footnote{Optionally $\phi_{\tbox{prt}}$
is determined by ${ \Gamma(\phi) \sim \hbar/\tau_{\tbox{cl}} }$. 
It should be distinguished from the border of the first order 
perturbative regime which is determined by ${\Gamma(\phi) \sim \Delta}$, 
leading to $\phi_{\tbox{FOPT}} \sim \phi_{\tbox{prt}}/\sqrt{b}$ 
where ${b=(\hbar/\tau_{\tbox{cl}})/\Delta \gg 1}$. 
In other words $\phi_{\tbox{FOPT}}$ is the 
perturbation which is needed to mix neighboring levels.} )
is ${\phi_{\tbox{prt}} = L_x / (v_E \tau_{\tbox{cl}}) \sim 1}$.

\begin{figure}
\includegraphics[clip,width=1.0\hsize]{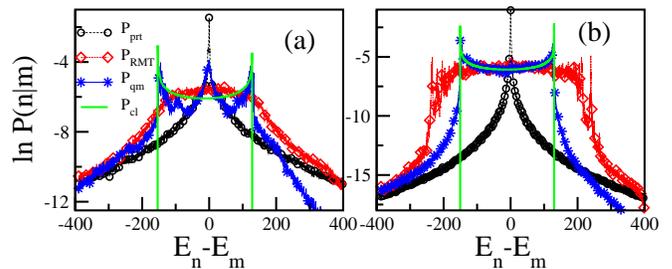}
\caption{
(a) The LDOS kernel $P(n|m)$ for $\phi=31.4$, where $\ell=100$.  
(b) The same parameters but $\ell=1$. In panel (a) 
we observe co-existence of perturbative and SC structures
while in panel (b) we witness detailed QCC.}
\end{figure}


What happens to $P(r)$ in practice?  If we take the Wigner RMT model 
as an inspiration, we expect to have at $\phi\sim\phi_{\tbox{prt}}$ 
a simple crossover from a $P_{\tbox{prt}}$ line-shape to a $P_{\tbox{cl}}$ line-shape. 
The latter is regarded as the semiclassical analogue 
of the (artificial) semicircle line shape. 
Indeed for the smooth billiard ($\ell=1$) we have verified that 
this naive expectation is realized \cite{MCK04}.
But for the rough billiard ($\ell=100$) we witness a more 
complicated scenario. In Fig.~2(b,c) we show the LDOS for $\phi<\phi_{\tbox{prt}}$, 
where it (still) agrees quite well with $P_{\tbox{prt}}$. 
In Fig.~3 we show the LDOS for $\phi>\phi_{\tbox{prt}}$, 
where we would naively expect agreement with $P_{\tbox{cl}}$. 
Rather we witness a three peak structure, where the $r\sim0$ peak 
is of perturbative nature, while the other are the fingerprint of semiclassics.
For sake of comparison we show the corresponding results 
for a smooth billiard ($\ell=1$) and otherwise the same parameters.
There we have detailed QCC as is naively expected.
The co-existence of perturbative and semiclassical features  
persists within an intermediate regime of $\phi$ values, 
to which we refer as the ``twilight zone".


Before we adopt a phase space picture in order to explain 
the above observations, we would like to verify that indeed random matrix 
modeling does not lead to a similar effect: 
After all the standard Wigner model, 
that gives rise to a simple crossover from a Lorentzian 
to a semicircle line shape, assumes a simple banded matrix,  
which is {\em not} the case in our model.  As argued above 
the matrix elements of $\bm{B}$ decay as $1/|n-m|^2$ from 
the diagonal. This implies that $P_{\tbox{prt}}(r)$ is in fact not 
a Lorentzian, and also may imply that the crossover 
to the non-perturbative regime is more complicated. 
In order to resolve this subtlety we have taken a randomized 
version of the Hamiltonian  ${\cal H} = \bm{E} + \phi \bm{B}$.  
Namely, we have randomized the signs of the off-diagonal 
elements of the $\bm{B}$ matrix. Thus we get an RMT model 
with the same band profile as in the physical model.
This means that $P_{\tbox{prt}}$ is the same for both models 
(the physical and the randomized), but still they can 
differ in the non-perturbative regime. Indeed, 
looking at the LDOS of the randomized model we observe 
that the semiclassical features are absent:  
$P_{\tbox{RMT}}(r)$ unlike $P(r)$ exhibits a simple crossover 
from perturbative to non-perturbative lineshape.


\begin{figure}
\includegraphics[clip,width=0.9\hsize]{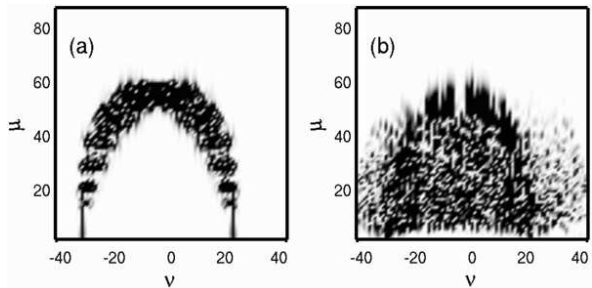}
\caption{
The probability distribution $|\langle \nu,\mu |n \rangle|^2$ for 
(a) The $n=2423$ eigenstate of the smooth ($\ell=1$) billiard; 
(b) The $n=1000$ eigenstate of the rough ($\ell=100$) billiard. 
Note that this is essentially the $(p_x,p_y)$ momentum distribution. 
The state in panel (a), unlike the state in panel (b), 
is a typical semiclassical state. Namely it is well concentrated 
on the energy shell.}
\end{figure}

In what follows we would like to argue that the structure of $P(r)$, 
both perturbative and non-perturbative components, 
can be explained using a {\em phase space picture}. 
[For phrasing purpose we find the ``Wigner function language'' 
most convenient, still the reader should notice that 
we do not need or use this representation in practice]. 
We recall the $P(n|m)$ is determined by the overlap of two Wigner functions.
In the present context the Wigner functions $\rho^{(n)}$ are supported 
by shifted circles $(p_x-(\phi/(2\pi))^2+p_y^2 = 2\mathsf{m}E_n$. 
We are looking for their overlap with a reference Wigner function which 
is supported by the circle $p_x^2+p_y^2 = 2\mathsf{m}E_m$. 
The question is whether the overlaps of the Wigner functions 
$\rho^{(n)}$ and $\rho^{(m)}$ can be approximated by a 
classical calculation, and under what circumstances we 
need perturbation theory.

Generically the Wigner function has a transverse Airy-type 
structure. If the ``thickness" of the Wigner function is 
much smaller compared with the separation $|E_n{-}E_m|$ of the energy surfaces 
then we can trust the semiclassical approximation. 
This will always be the case if $\hbar$ is small enough, 
or equivalently if we can make $\phi$ large enough. In such case 
the dominant contribution comes from the intersection of 
the energy surfaces, which is the phase space analogue of stationary 
phase approximation. 
The other extreme is the case where the ``thickness" of Wigner function 
is larger compared with the separation of the energy surfaces 
(namely $\delta E_{\tbox{cl}}(\phi) < \hbar/\tau_{\tbox{cl}}$). 
Then the contribution to the overlap 
comes ``collectively" from all the regions of 
the Wigner (quasi) distribution, not just from the intersections. 
In such case we expect perturbation theory to work.

The above reasoning assumes that the wavefunction is 
concentrated in an ergodic-like fashion in the vicinity of 
the energy surface. This is known as ``Berry conjecture" \cite {berry}. 
In case of billiards it implies that the wavefunction 
looks like a random superposition 
of plane waves with \mbox{$|p|=(2\mathsf{m}E)^{1/2}$}. 
We find (see Fig.~4) that this does not hold in case of a rough billiard 
(unless $\hbar$ were extremely small, 
so as to make the De-Broglie wavelength very short). 
Namely, in the case of a rough billiard 
there are eigenstates that have a lot of weight 
in the region \mbox{$|p|<(2\mathsf{m}E)^{1/2}$}.   
Consequently there are both semiclassical and non-semiclassical overlaps. 
Specifically, if we have non-semiclassical wavefunctions, 
and $|E_n-E_m|\sim 0$, then the {\em collective} contribution 
dominates, which give rise to the perturbative-like peak in the LDOS.

Our findings apply to systems, such as the rough billiard,  
where there is an additional (large) {\em non-universal} energy 
scale $\delta E_{\tbox{NU}}$. This is defined as an energy scale 
which is {\em not related} to the bandprofile, 
and hence does not emerge in the RMT modeling. 
Hence in general there is a distinct twilight regime  
\mbox{$\hbar/\tau_{\tbox{cl}} < \delta E_{\tbox{cl}}(\phi) < \delta E_{\tbox{NU}}$}, 
which is neither ``perturbative" nor ``semiclassical". 
[In our numerics $\ell{=}100$ is so large that $\delta E_{\tbox{NU}}{\sim} E$.]


{\bf Summary:} We have analyzed the parametric evolution of the 
eigenstates of an Aharonov-Bohm cylindrical billiard, 
as the flux is changed. For the first time the full 
crossover from the perturbative to the non-perturbative 
regime is demonstrated. Random matrix theory suggests 
a {\em simple} crossover. Instead, we discover an intermediate 
twilight regime where perturbative and semiclassical features 
co-exist. This can be understood by adopting a phase space picture,  
and taking into account the inapplicability of the Berry conjecture 
regarding the semiclassical structure of the wavefunctions.  

 
This research was supported by a grant from the GIF, 
the German-Israeli Foundation for Scientific Research and Development, 
and by the Israel Science Foundation (grant No.11/02).



\end{document}